# Correlation of conductivity and angle integrated valence band photoemission characteristics in single crystal iron perovskites for 300 K < T < 800 K: Comparison of surface and bulk sensitive methods

A. Braun[1a], B.S. Mun[2,3], Y. Sun[4], Z. Liu[2], R. Mäder[1], S. Erat[1,5], X. Zhang[6,7], S.S. Mao[6,7], E. Pomjakushina[8], K. Conder[8], L. J. Gauckler[5], T. Graule[1,9]

[1] *Department of Modern Materials and Surfaces*
*EMPA, Swiss Federal Laboratories for Materials Testing & Research*
*CH – 8600 Dübendorf, Switzerland*

[2] *Advanced Light Source, Ernest Orlando Lawrence Berkeley National Laboratory*
*Berkeley CA 94720, USA*

[3] *Department of Physics, Hanyang University, Ansan, Republic of Korea*

[4] *Stanford Synchrotron Radiation Lightsource, Menlo Park CA 94025, USA*

---

[a] *Corresponding author. Phone +41 (0)44 823 4850, Fax +41 (0)44 823 4150, email: artur.braun@alumni.ethz.ch*

[5]*Department for Nonmetallic Inorganic Materials, ETH Zürich,*

*Swiss Federal Institute of Technology*

*CH-8037 Zürich, Switzerland*

[6]*Department of Mechanical Engineering, University of California Berkeley*

*Berkeley CA 94720, USA*

[7]*Environmental Energy Technologies Division, Lawrence Berkeley National Laboratory*

*Berkeley CA 94720, USA*

[8]*Laboratory for Developments and Methods, Paul Scherrer Institut*

*CH – 5232 Villigen PSI, Switzerland*

[9]*Technische Universität Bergakademie Freiberg, D-09596 Freiberg, Germany*

**Abstract**

A single crystal monolith of $La_{0.9}Sr_{0.1}FeO_3$ and thin pulsed laser deposited film of $La_{0.8}Sr_{0.2}Fe_{0.8}Ni_{0.2}O_3$ were subject to angle integrated valence band photoemission spectroscopy in ultra high vacuum and conductivity experiments in ambient air at temperatures from 300 K to 800 K. Except for several sputtering and annealing cycles, the specimen were not prepared in-situ.. Peculiar changes in the temperature dependent, bulk representative conductivity profile as a result of reversible phase transitions, and irreversible chemical changes are semi-quantitatively reflected by the intensity variation in the more surface representative valence band spectra near the Fermi energy. X-ray photoelectron diffraction images reflect the symmetry as expected from bulk iron perovskites. The correlation of spectral details in the valence band photoemission spectra (VB PES) and details of the conductivity during temperature variation suggest that valuable information on electronic structure and transport properties of complex materials may be obtained without in-situ preparation.

# 1. Introduction

The electric properties of condensed matter are particularly relevant in materials that are used in the energy sector. Depending on which components they comprise, they shall have insulating, conducting or semi-conducting properties. The requirement for specific *electronic* transport properties is sometimes paralleled by requirement for specific *ionic* transport properties, in particular in electrochemical energy conversion and storage devices like fuel cells [Sun, Maguire], batteries, and capacitors. While batteries, for example, are a daily life commodity, their structural and chemical complexity is not always obvious. The same holds for fuel cells.

Electric transport properties of materials are to a wide extent determined by their electronic structure, which can be well studied with x-ray and electron spectroscopy [Imada 1998]. Photoemission spectroscopy has contributed significantly to understanding the fundamental principles of conduction and band structure in metals and semiconductors [Damascelli 2003]. This understanding was enhanced by the ability to perform angle resolved photoemission spectroscopy (ARPES) studies on single crystals, and experimental data could validate computational theoretical models [Imada 1998].

There are two basic shortcomings with such PES experiments: 1) they are highly surface sensitive, whereas the electronic conductivity may be important as a bulk property, such as in electrodes and conductors; 2) the PES experiments are carried out at very low temperature or, at best, ambient temperature, whereas in daily life there may be situations where conductivity is important at elevated temperatures, i.e. significantly above 300 K, for instance in gas sensors and high temperature fuel cells.

We present here a challenge to the conventional applications of angle resolved photoemission or ultraviolet spectroscopy, which should be interesting to the community.

The ceramic solid oxide or solid electrolyte fuel cell (SOFC) operates at temperatures as high as 1000°C. Due to technological progress in the past years, the temperatures could be lowered to 800°C, and there is hope that further improvement of technology allows to build systems that work at intermediate temperatures (IT) of 400°C – 500°C (IT-SOFC). The high temperatures are necessary in order to activate the oxygen ion or proton conductivity in the ceramic electrolyte [Braun 2009]. The other components, i.e. ceramic cathodes and anodes, and the metallic interconnectors face the same high temperature. Current SOFC cathodes are comprised of LaSrMn-oxide, a mixed electronic ionic conducting (MIEC) perovskite with $ABO_3$ structure, a material well known to condensed matter physicists as *strongly correlated electron system*.

An alternative for MIEC at intermediate temperatures is LaSrFeNi-oxide, a small polaron conductor which has a conductivity maximum in the IT-SOFC operation range [Swierczek 2006]. We have systematically studied single crystalline LaSrFe and LaSrFeNi oxide with x-ray photoelectron spectroscopy at excitation energies of 60eV, 120eV and 450 eV at temperatures up to 600°C. LSF and LSFN are 3-dimensional $ABO_3$ type perovskites and cannot be cleaved easily. They were prepared ex-situ and annealed, or Ar sputtered and annealed before recording spectra.

## 2. Experimental

$La_{0.9}Sr_{0.1}FeO_3$ (LS10F) powder was synthesized in a solid state thermal reaction by mixing $La(OH)_3$, $SrCO_3$ and $Fe_2O_3$ in the stoichiometric amounts, heated up to 1473

K with 5 K/min and held for 10 h at temperature, followed by a cooling rate of 5 K/min. Phase purity was confirmed by x-ray diffraction. Crystallographic and electronic structure data of LSF are available elsewhere [Dunn, Fossdal, Braun 2008, Haas 2009].

To obtain the single crystal, the powder was pressed in a cold isostatic press (Vitec) at 2000 bar to a rod of 8 mm in diameter and pre-sintered at 1573 K for 4 h. The seed and feed rods were pressed from the powder at 4 kbar in a hydrostatic press and sintered at 1523 K (12h) in air. The single crystal was grown in an optical floating zone furnace (CSC Japan), powered with 1000 Watt halogen tubes at a growth rate of 1.5 mm/h in Ar mixed with 1% $O_2$ atmosphere at 4 bar, with 15 rotations per minute. From the single crystal slab, a 1 mm thick slice in the [111] orientation was cut and high-quality surface finished (CRYSTEC, Düren, Germany), Figure 1-a.

A 3 cm long part of the slab was used to measure the DC conductivity as a function of temperature by the four-point method in a custom made furnace with computer controlled atmosphere and temperature.

The monolithic single crystal with [111] orientation was subject to angle integrated x-ray photoemission spectroscopy (PES) with 450 meV excitation energy at beamline 9.3.2 at the Advanced Light Source in Berkeley, California, in an UHV recipient with $10^{-10}$ Torr base pressure. Prior to the PES, the sample was subject to $Ar^+$ ion bombardment for 30 minutes at 500 eV, a subsequent heating cycle to 500 K and back to ambient temperature, and a scan for residual carbon on the surface. This procedure was repeated two times, after which no carbon signal was detected on the crystal surface. Survey scans were made from -10 to +150 eV binding energy $E_B$ at temperatures from 323 K to 723 K in steps of 50 K. The survey scans, shown in Figure 2, were normalized relative to the La peak.

A 175 nm thick film with high quality, black luster surface was pulsed laser deposited on a single crystal SrTiO$_3$ (110) substrate (lattice constant a=0.3905 nm, Crystec, Berlin, Germany) from a (La$_{0.8}$Sr$_{0.2}$)$_{0.95}$Ni$_{0.2}$Fe$_{0.8}$O$_3$ (LSFN) feed target for 45 minutes at 773 K and 10 mTorr oxygen base pressure, and then in-situ annealed with 250 Torr oxygen (Figure 1-b). The feed target was prepared in the same manner like that for the melt grown LSF single crystal [Braun 2009a].

For this LSFN film, PES spectra were collected at BL 8-1 at Stanford Synchrotron Radiation Lightsource in ultrahigh vacuum of 2.1·10$^{-9}$ Torr, after careful surface cleaning at 623 K and monitoring volatilization of surface adsorbates with survey scans. After cooling to 323 K, the film was heated to 831 K while spectra were recorded at 60 eV and 120 eV for various T. Arrived at 831 K, the film was flash heated to 1023 K for 5 minutes and returned to 828 K. During cooling, spectra were recorded until T = 344 K was reached. After that, the film had changed its black opaque color to then reddish, translucent color.

The 4-point DC conductivity as a function of T was obtained with a computer controlled Keithley 2400 sourcemeter and a Keithley 2700 multimeter in a muffle furnace under ambient air pressure.

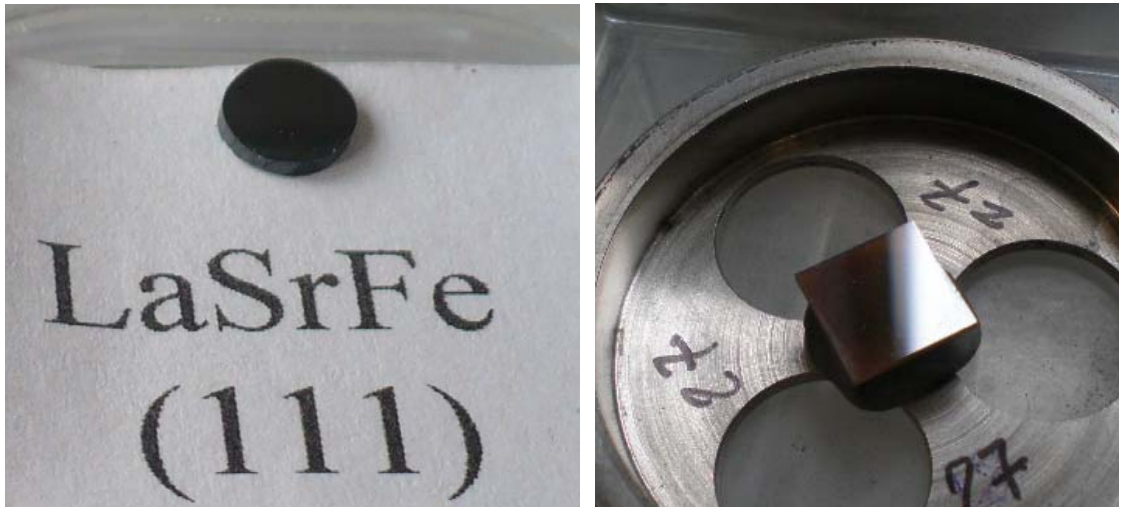

**Figure 1: -a)** Monolithic single crystal LS10F disk in (111) orientation. **–b)** 175 nm PLD film of LSFN on SrTiO$_3$(110).

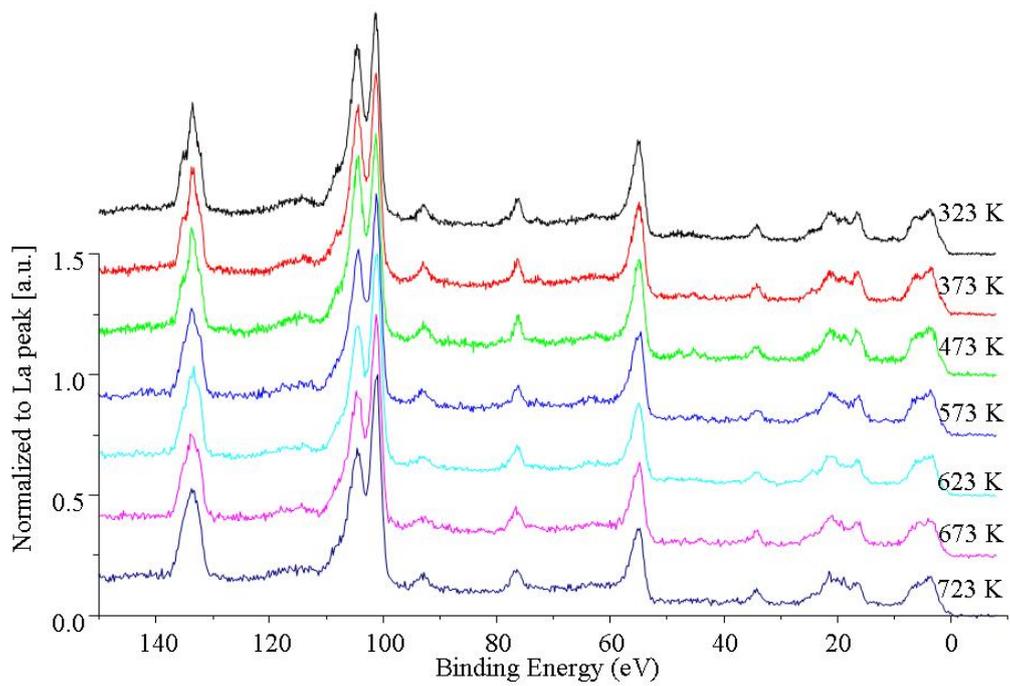

**Figure 2:** Survey scan for LS10F at different temperatures 323 K – 723 K.

## 3. Results and discussion

Lanthanum strontium iron oxides (LSF) have a distorted perovskite structure [Vracar 2007, Blasco 2008]. The parent compound $LaFeO_3$ has orthorhombic symmetry with four distorted pseudo-cells in the space group Pbnm, and is an antiferromagnetic charge transfer insulator with 2 eV charge gap energy. Each Fe atom has its spin antiparallel to its six nearest neighboring Fe atoms. The Néel temperature is 738 K and drops, when Sr is added. Parasitic ferro- and ferrimagnetism are found, including permanent magnetization, when the Sr concentration is kept small [Grenier 1984, Dunn 1994]. The heterovalent substitution of $La^{3+}$ in the insulating $LaFeO_3$ with $Sr^{2+}$ (hole doping) creates electron hole states with substantial O(2p) character near the Fermi level [Sarma 1994], whereas subsitution on the B-site with Ni creates d-type holes, which goes along with a significantly increased conductivity [Erat 2009]. The electric conductivity of $La_{(1-x)}Sr_xFeO_3$ increases with increasing Sr doping to an extent, that $SrFeO_3$ shows typical metal behavior [Imada 1998]. $SrFeO_3$ is an antiferromagnetic metallic conductor with temperature dependent conductivity [Park 1999] and Neel temperature of 134 K [Imada 1998].

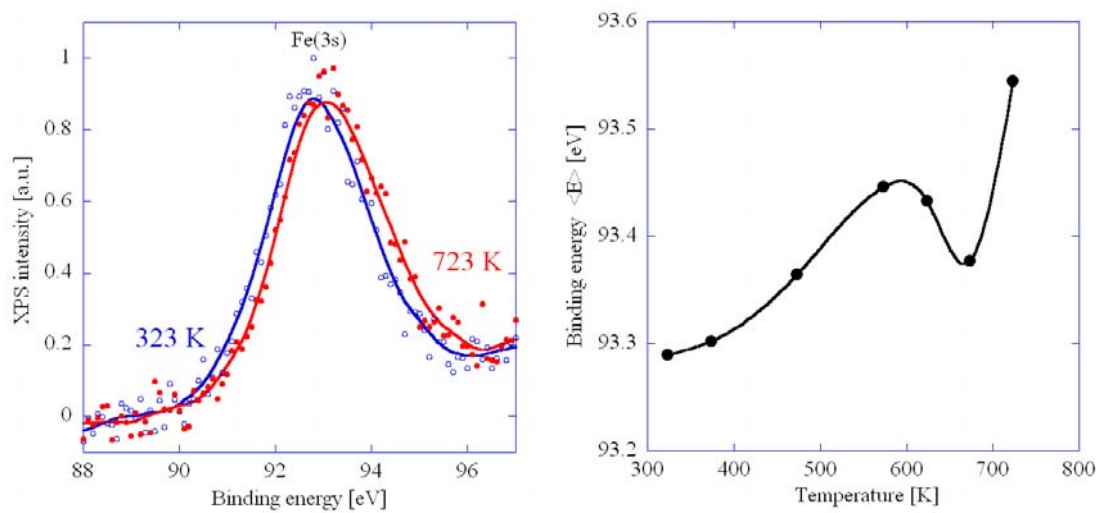

**Figure 3:** Fe(3s) core level emission spectra for 323K and 723K show a chemical shift of around 0.3 eV, suggesting a reduction of the Fe with increasing T. The first moment <E> of the binding energy is plotted for all measured T in the right panel.

Figure 3 shows the Fe(3s) core level spectra of LSF from 323 K to 723 K. Both spectra are shifted on the energy scale. In order to quantify the energy position for all spectra recorded at temperatures in the range 323 K < T < 723 K, we determined the first statistical moments <E>;

$$\langle E \rangle = \frac{\int_{88eV}^{97eV} I(E) \cdot E \cdot dE}{\int_{88eV}^{97eV} I(E) \cdot dE}$$

[Braun 2002], which are plotted on the right panel in Figure 3 for all temperatures. Obviously, the spectra are shifting towards higher binding energies, indicating a reduction of the Fe upon annealing. Interestingly, at about 573 K the first moment <E> decreased again, until a minimum at around 675 K, and a subsequent steep increase of the <E> with increasing temperature. The overall shift in binding energy across the temperature range probed is about 0.25 eV. The nominal average oxidation state of Fe in $La_{0.9}Sr_{0.1}FeO_3$ is $Fe^{3.1+}$. Thus, nominal we have 10% $Fe^{4+}$ and 90% $Fe^{3+}$. LSF compounds with Sr doping as little as 10% are virtually stoichiometric.

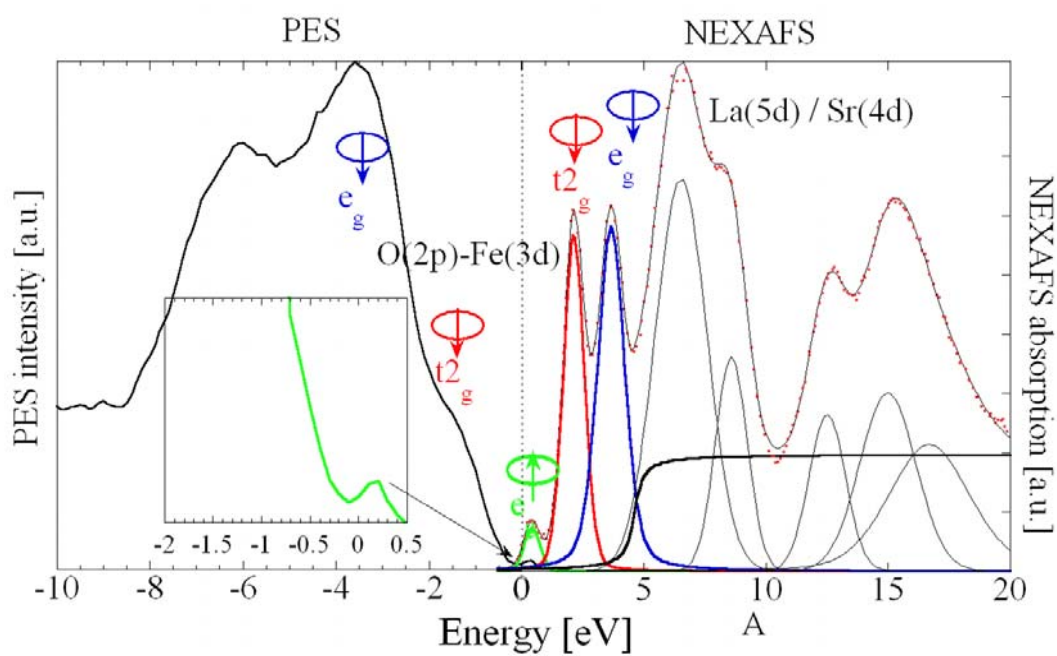

**Figure 4:** left – VB PES spectrum of LS10F. Inset shows magnified doped hole resonance. right – oxygen NEXAFS spectrum of LS10F, deconvoluted by Voigt functions and arctan function.

Figure 4 shows the photoemission spectrum (PES) in the valence band (VB) region measured at 323 K. The peak intensity at around -3.2 eV is assigned to a spin down $t_{2g}$ ($t_{2g} \downarrow$) transition. For the ambient temperature spectrum, there is a shoulder clearly visible at around -1.4 eV, which is assigned to a spin down $e_g$ ($e_g \downarrow$) transition [Wadati 2006]. A very small intensity peak is noticeable at about the Fermi energy $E_F$; this state is from electron holes which are doped as a result of Sr substitution on the A-site (inset in the left panel in Figure 4). These states are paralleled in the oxygen NEXAFS spectrum in the right panel in Figure 4. Same like the VB PES spectra, the oxygen NEXAFS spectra show hybridized states from Fe(3d) and O(2p) orbitals. We could recently show that the intensity ratio of spin up to spin down of these three states in the valences band, as probed with oxygen NEXAFS, scales exponentially with the electric conductivity in A-site and B-site site substituted iron perovskites [Braun 2009b].

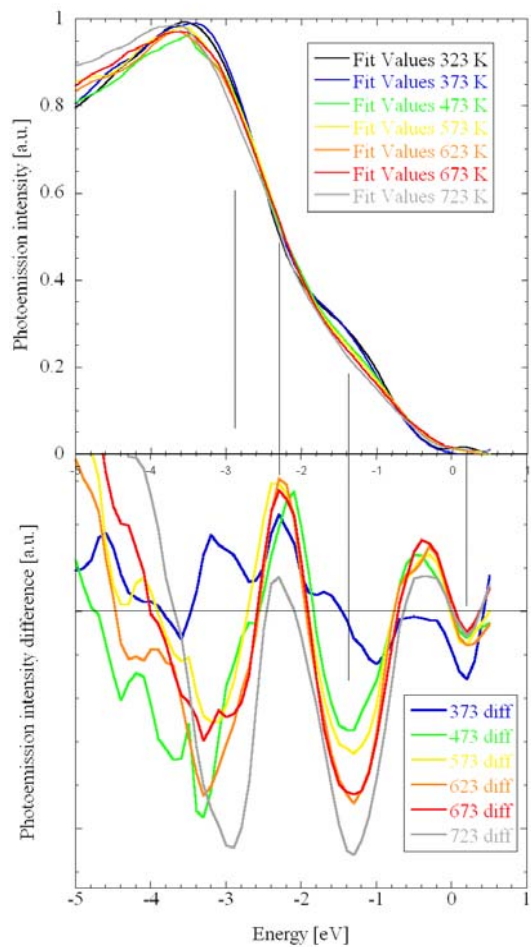

**Figure 5:** top - temperature dependent VB PES spectra for 323 K < T < 723 K of LS10F. bottom – difference of the spectra for all T with respect to the spectrum measured at 323 K.

Close inspection of the binding energy range larger than -2 eV shows that the intensity of this shoulder decreases with increasing temperature, whereas this intensity is redistributed towards higher binding energy. In fact, the tails intersect at around -0.75 eV, and the spectral intensity increases at $E_B > -0.75$ eV with increasing temperature.

The bottom panel of Figure 5 shows the difference between the high temperature spectra and the spectrum recorded at 323 K. The general trend is that this intensity is increasing with increasing temperature. The spectrum at 373 K shows the lowest PES intensity near $E_F$, even lower than so at 323 K.

Since we are looking at temperature activated properties, it is worthwhile to consider corresponding quantities in Arrhenius representation, or at least versus the inverse temperature. Hence, in Figure 6 (top panel), we have plotted the PES intensities near $E_F$ versus the absolute temperature. The data points are connected by a cubic spline to guide the eye. The temperatures with the three highest spectral intensities are 323 K, 573 K, and 723 K.

The electric conductivity versus T is shown in Figure 6 (bottom panel) in Arrhenius representation. We notice here three details: i) the crossover from semiconductor type polaron conductivity to metallic conductivity at around 700 K; ii) a small jump in the conductivity at 357 K, and iii) a change in the activation energy $E_a$ at 573 K. The different ranges are highlighted by linear fits, which basically correspond to the activation energies $E_a$. These fitting lines are extended in order to visualize the temperature of the onset of the changes, i.e. 357 K, 573 k, and 700 K.

We notice that at around 573 K, where the activation energy changes from 317 meV to 332 meV, also the first statistical moment $<E_B>$ of the Fe(3s) XPS peaks shows a change from reductive to oxidative behavior. This becomes particularly obvious when $<E_B>$ is plotted versus reciprocal temperature (second panel from bottom in Figure 6). The oxidative trend returns to the reductive trend at 700 K, where the conductivity mechanism changes from semiconducting polaronic to metal-like, and where the PES intensity variation shows already the same correlation. Differential scanning calorimetry (DSC) data, also plotted versus the inverse temperature, show noticeable alterations at about the same temperatures (second panel from top in Figure 6): A relatively constant increase from 300 K to about 573 K, and then a steep decrease towards higher temperatures. Maxima at about 573 and 700 K coincide with the change in activation energy and change of conductivity mechanism.

The small jump in the conductivity at 357 K and the corresponding minimum in the PES height intensity near $E_F$ cannot be identified, however, in the DSC curve or Fe(3s) core level spectra.

LS10F undergoes (at least two) phase transitions from cubic to orthorhombic and then rhombohedral during annealing [Braun APL 2008]. It is therefore not surprising that structural changes are accompanied by changes in the electronic structure, which reflect on the transport properties. Even the very minute changes in the conductivity like the one at 357 K may have their molecular origin in barely noticeable structural changes.

The aforementioned property changes are measured mainly at bulk sensitivity, whereas the spectroscopy data are very surface sensitive, given the excitation energy of 450 eV and the small escape depth of the photoelectrons. But the good correlation of VB PES information and the conductivity and calorimetry data suggests that we are indeed probing to some extent the bulk properties with the PES method, or in other words, that the surface of the repeatedly annealed and $Ar^+$ sputtered LS10F represents to a significant extent its bulk properties. In order to check and validate in how far the crystallographic structure of the LS10F surface matches that of the bulk LS10F, we subjected the single crystal to x-ray photoelectron diffraction (XPD).

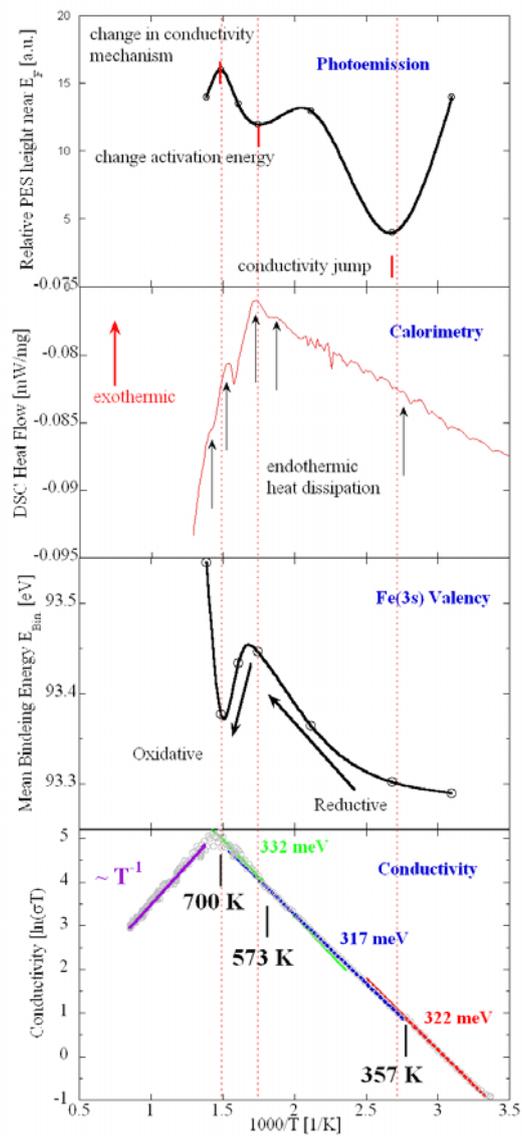

**Figure 6:** Correlation of VB PES intensity near EF (top), electric conductivity (bottom), and chemical state of the Fe and the calorimetric data of LS10F single crystal during annealing.

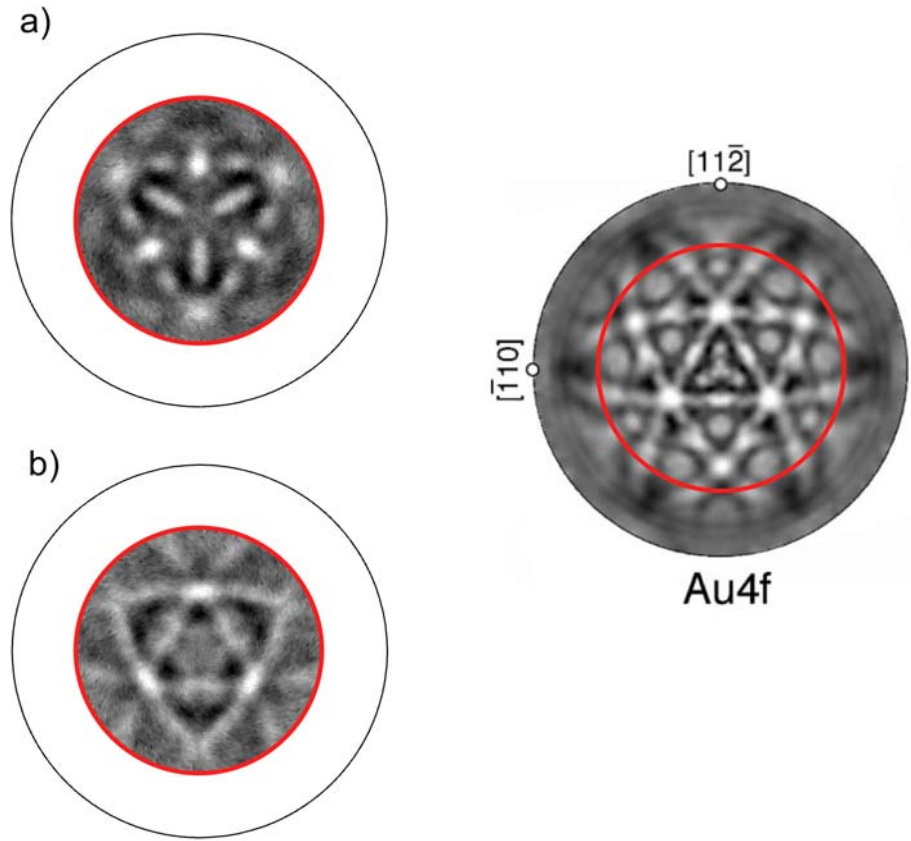

**Figure 7:** X-ray photoelectron diffraction patterns of LS10F from the La 3d(5/2) line (left) and, for comparison, the Au4f line of Au(111).

Figure 7 shows stereographic projections of the angle resolved photoemission intensity of the La 3d(5/2) line at 833.2 eV binding energy and the O 1s line at 529 eV binding energy. The x-ray photoelectron diffraction (XPD) patterns were measured azimuthal from 0° to 65° in and averaged over symmetry multiplicity 3, i.e. for 0°+α, 120°+α, and 240°+α. In addition, φ averaging was applied, i.e. for constant azimuth the averaged intensity of all polar angles 0°-360° was subtracted. For comparison, the XPD pattern of Au(111) with bcc structure is shown. The red circle shows the 65° azimuth. Thus, by comparison, the bcc structure for the $La_{0.9}Sr_{0.1}FeO_3$ becomes immediately obvious.

Systems where this issue of bulk and surface properties have been tested extensively in recent years, is for instance LaSrMn-oxide, which shows the colossal magnetoresistance effect. A supposedly safe approach for ARPES appears to be to prepare films in-situ with PLD or molecular beam epitaxy (MBE) at the (synchrotron) beamline prior to (AR-)PES measurements [Shi 2004]. Kumigashira et al. [Kumigashira 2004] used in-situ MBE and showed that no subsequent surface cleaning was necessary to perform ARPES experiments. Lev et al. [Lev 2004] used a melt grown LSM single crystal and cleaved it in-situ in the UHV recipient, and they did not report any subsequent cleaning procedure. Kobayashi et al. [Kobayashi 2007] studied SrFeMo-oxide ceramic samples which they repeatedly scraped in-situ with a diamond file.

The left panel in Figure 8 shows the PES spectra of LSFN in the VB region for $0 < E_B < 25$ eV and 295 k < T < 623 K, obtained at 120 eV excitation energy. We immediately identify three classes of spectra depending on the temperature at which they were recorded. The five spectra recorded at 295 K < T < 373 K have the overall lowest intensity, with intensity maxima at about -6 eV, -17 eV, and -19.3 eV. In the VB region near $E_F$, we identify in addition to the maximum at -6 eV also two shoulders at about -3.5 eV and -2 eV.

The next class of spectra is those recorded at 473 K, 533 K and 683 K. Their overall intensity is by a factor of three to four higher than the previous ones. The intensity maxima at -17 eV and -19.3 eV from the previous class are recovered, but the previous shoulder structure at -3.5 eV has now developed into a peak. This peak is the leading feature in the next class of spectra, recorded at 623 K, which is also by a factor of 10 higher in intensity than the class measured in the lowest temperature range. Overall, we notice that between 373 K and 473 K, and between 583 K and 623 K, processes on the surface of the LSFN must occur that cause this significant alteration of the overall spectral intensity. Note, however, that the spectral details, i.e. peaks and shoulders, are present in all spectra at the same energies. This suggests that the sample as such is not so much affected by the temperature change that it might appear at a first glance. The most likely processes are the volatization of surface adsorbates like $H_2O$ and $CO_2$.

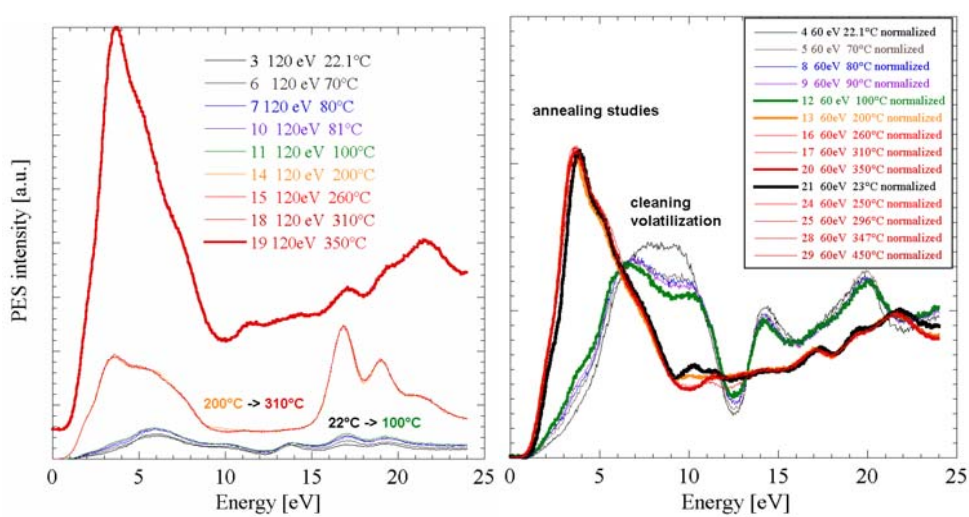

**Figure 8**: left - VB PES spectra of LSFN at 120 eV excitation energy for 295 K<T<623 K. right - VB PES spectra for LSFN at 60 eV excitation energy for 295 K<T<723 K. The numerals before the energy denote the chronological scan number.

The right panel in Figure 8 shows the same temperature study, but at 60 eV excitation energy. The spectra are plotted and also normalized in a way that they show two classes of spectra. The spectra that tend with the maximum to the higher binding energy are those from the cleaning and volatization study. After 200°C (473 K) sample temperature was reached, the spectra have a general shape that remains the same up to 350°C (623 K), even after returning to 23°C (296 K), and subsequent annealing to 450°C (723 K).

At 723 K, the thin film LSFN sample was flash annealed to 828 K for few minutes, and then slowly returned to ambient temperature. The VB PES spectra recorded during this temperature history show a substantial change after the flash annealing. Visual inspection shows that the PES intensity near $E_F$ is high at 831 K and decreases during cooling to 344 K. Quantitative analysis shows that that the intensity actually oscillates between 600 K and 500 K, in the same way like the electric conductivity of this film [Braun 2009 APL paper].

Figure 9 shows the PES intensity near $E_F$ for all measured temperatures.

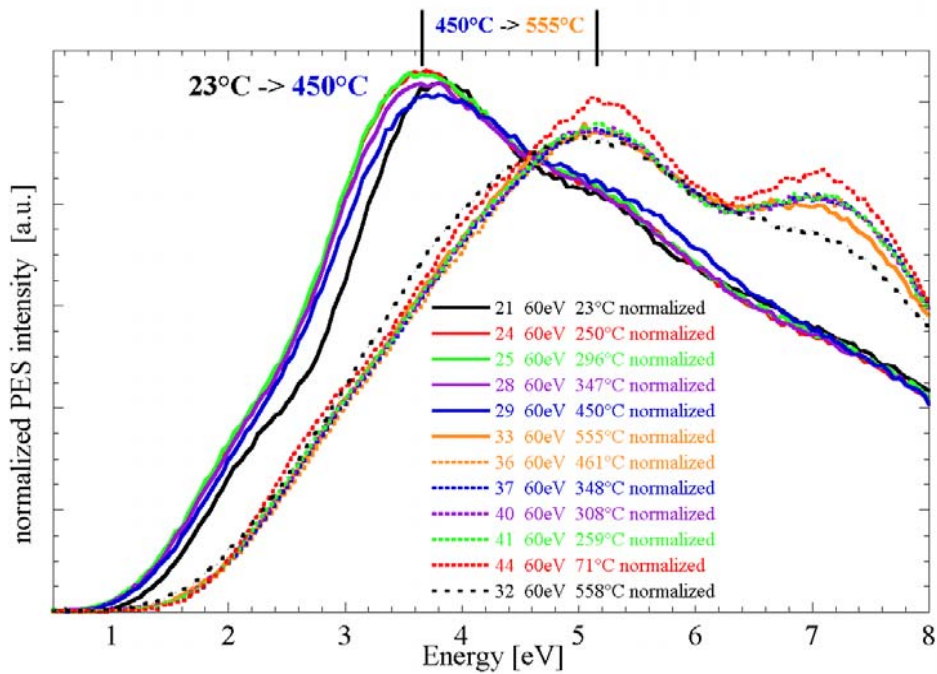

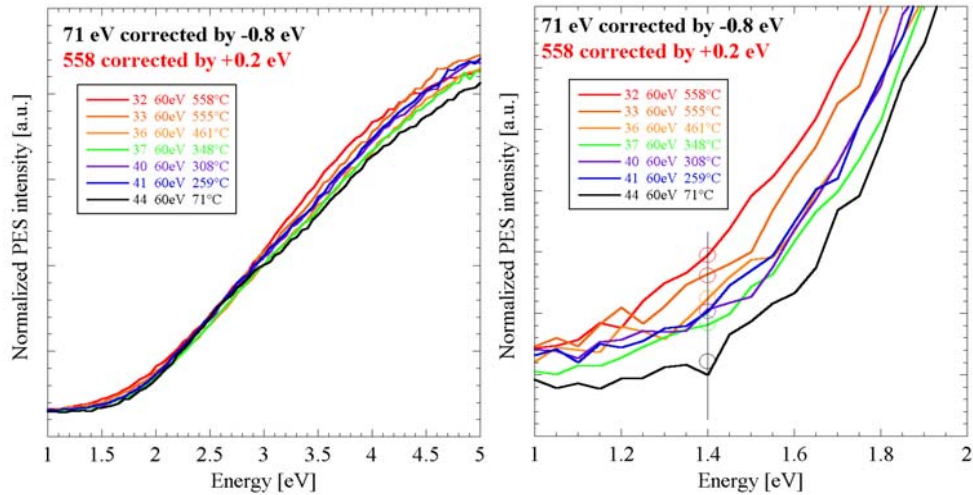

**Figure 9: -a)** VB PES spectra of LSFN at 120 eV excitation energy for 295 K<T<623 K. **–b)** VB PES spectra for LSFN at 60 eV excitation energy for 295 K<T<723 K. The numerals before the energy denote the chronological scan number.

**4. Summary**

Two examples where surface sensitive VB PES experiments and bulk sensitive conductivity experiments were combined show that surface cleaning procedures like annealing in UHV plus $Ar^+$ bombardment can provide surfaces for meaningful quantitative VB PES measurement which reflect the bulk properties like conductivity surprisingly accurate.

**Acknowledgment**

Financial support by the European Commission (MIRG # CT-2006-042095 and Real-SOFC # SES6-CT-2003-502612) and the Swiss National Science Foundation (SNF # 200021-116688) are acknowledged. The ALS is supported by the Director, Office of Science, Office of Basic Energy Sciences, of the U.S. Department of Energy under Contract No. DE-AC02-05CH11231. Part of this research was carried out at the Stanford Synchrotron Radiation Lightsource, a national user facility operated by